\documentclass[apl,article,superscriptaddress,reprint]{revtex4-1}

\usepackage{graphicx}
\usepackage{epsfig}
\usepackage{color}
\usepackage[hypertex,unicode=true]{hyperref}

\hypersetup{
        colorlinks=false,
        citecolor=blue,
        linkcolor=black,
        urlcolor=blue}

\newcommand{\un}   [1]{\ensuremath{\,\mathrm{#1}}}

\newcommand{\pderl}[2]{\ensuremath{\partial #1/\partial #2}}

\newcommand{\kint} {\ensuremath { \kappa_{\mathrm{int}} }}
\newcommand{\kext} {\ensuremath { \kappa_{\mathrm{ext}} }}
\newcommand{\Fint} {\ensuremath { \mathcal{F}_{\mathrm{int}} }}
\newcommand{\FSR}  {\ensuremath { \mathrm{FSR} }}
\newcommand{\Cout} {\ensuremath { C_{\mathrm{out}} }}

\begin{document}

\title{Characterization of optical quantum circuits using resonant phase shifts}

\author{M.~Poot}
\affiliation{Department of
Electrical Engineering, Yale University, New Haven, CT 06520,
USA}

\author{H.~X.~Tang}
\email{hong.tang@yale.edu} \affiliation{Department of
Electrical Engineering, Yale University, New Haven, CT 06520,
USA}

\date{\today}

\begin{abstract}
We demonstrate that important information about linear optical circuits can be obtained through the phase shift induced by integrated optical resonators. As a proof of principle, the phase of an unbalanced Mach-Zehnder interferometer is determined. Then the method is applied to a complex optical circuit designed for linear optical quantum computation. In this controlled-NOT gate with qubit initialization and tomography stages, the relative phases are determined as well as the coupling ratios of its directional couplers.
\end{abstract}

\maketitle
\newpage
Integrated optics is a promising candidate to implement quantum computing \cite{nielsen_book_quantum_computation, kok_RMP_LOQC}. In the linear-optics-quantum-computation framework, photonic circuits enable operations on qubits encoded with single photons \cite{knill_nature_quantum_computation_linear_optics}. High-fidelity operations require extensive design, fabrication, and characterization of its elements \cite{poot_OE_quantum_circuits}. For single components or simple circuit this can be done with careful calibration of transmissions. However, characterizing a complex as-fabricated photonic circuit solely through its input and output ports presents a major challenge. Previous work on integrated quantum circuits has employed, for example, imaging light scattered from the circuit \cite{holmgaard_JLT_plasmonic_directional_couplers, metcalf_natcomm_multiphoton_circuit}, measuring interference fringes while sweeping the phase between light at two inputs \cite{rahimi-keshari_OE_linear_charactization}, and iteratively optimizing the phases of all phase shifters \cite{carolan_science_universal_LOQC}. However, a flexible method that can be used with low-loss waveguides, and without having to electrically contact every device before measuring it, is extremely valuable. Also, in our vision \cite{poot_OE_quantum_circuits, poot_CLEO_integrated_quantum_optics} of fully-integrated quantum-optic devices with on-chip superconducting detectors and optomechanical phase shifters, the possibility to have in-situ calibration options is crucial. To this end, ring resonators that change the phase of light in a narrow wavelength range are incorporated in our circuits to enable full characterization of the important parameters. First the transfer function is discussed, and then the optimal properties of the rings are found. Then resonances are used to obtain the phase difference between the two paths of a Mach-Zehnder interferometer. Finally, with the help of phase shifts induced by optical ring resonators, a complex quantum-optics circuit is fully characterized.

\begin{figure}[tb]
\includegraphics{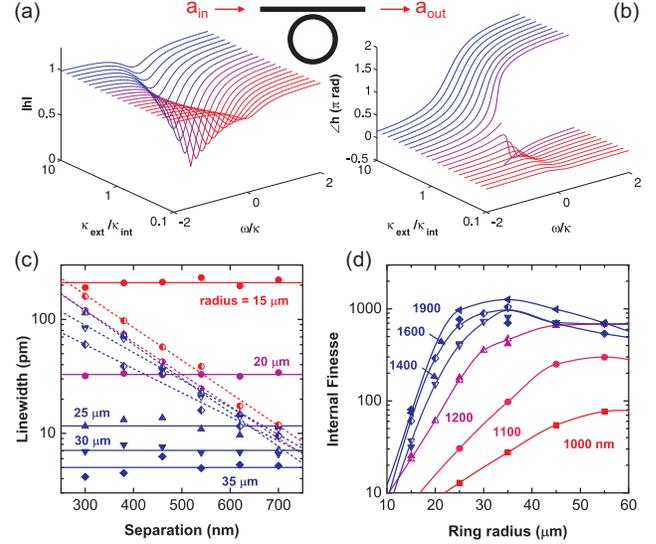}
\caption{(a) Amplitude and (b) phase of light traveling through a waveguide that is coupled to a ring resonator (see inset) versus the detuning from a resonance; different curves correspond to different coupling conditions.
(c) Experimental internal (filled symbols/solid lines) and coupling (half-open symbols/dashed lines) linewidths for 1600 nm wide SiN microrings versus the waveguide-ring separation. (d) Internal finesse for rings with different widths. The lines are guides to the eye; half-open and filled symbols correspond to two different chips.
\label{fig:rings}}
\end{figure}
A waveguide coupled to an optical resonator (Fig. \ref{fig:rings}) with a single isolated resonance has a transfer function $h$. The ratio between the output ($a_{\mathrm{out}}$) and input ($a_{\mathrm{in}}$) light fields is \cite{collett_PRA_squeezing_input_output, bogaerts_LPR_ring_review}:
\begin{equation}
h(\omega) \equiv \frac{a_{\mathrm{out}}}{a_{\mathrm{in}}}= 1-\frac{\kint} {\kappa/2+i(\omega - \omega_0)}. \label{eq:transfer_ring}
\end{equation}
The total linewidth $\kappa = \kint + \kext$ is the sum of the internal ($\kint$) and external ($\kext$) damping rates. The former is due to loss in the resonator itself, whereas the latter is due to the coupling to the waveguide. $\omega/2\pi$ is the frequency; the resonance is at $\omega_0/2\pi$. Plots of the magnitude and phase of $h$ are shown in Fig. \ref{fig:rings}(a) and (b). For weakly-coupled resonances ($\kext \ll \kint$) the transmission $|h|$ shows a shallow dip and the phase $\angle h$ displays a small change. For increasing $\kext$, the dip becomes deeper and the change in the phase grows. Exactly at critical coupling $\kext = \kint$, $|h(\omega_0)|$ is zero and the phase is discontinuous. When further increasing $\kext$, the dip becomes shallower again, but the phase change remains $2\pi$. Thus, when the wavelength is swept over any overcoupled resonance ($\kext > \kint$) the phase of $a_{\mathrm{out}}$ is varied over all possible values. This is an important point that enables efficient characterizing of photonic circuits, in particular for our linear-quantum-optics platform \cite{poot_OE_quantum_circuits}. However, note again that the rings will be used to extract information about the circuit, in particular to \emph{determine} the optical phase differences; external phase shifters will be added in a later stage to \emph{control} the phases \cite{poot_CLEO_integrated_quantum_optics}.

Resonances are thus very useful to characterize a photonic circuit, but they should have a minimal impact on its operation. Hence it is important that they occupy only a minimal part of the spectrum, or appear at wavelengths that are not used for circuit operation. To study the dependence of the resonator properties on the ring geometry (for details of the fabrication, see Ref. \cite{poot_OE_quantum_circuits}), the transmission of waveguides coupled to rings has been measured. The observed resonances are fitted \cite{poot_OE_quantum_circuits} to extract $\kint$, $\kext$, as well as the free spectral range $\FSR$. Figure \ref{fig:rings}(c) shows the dependence of $\kint$ and $\kext$ on the separation $s$ between the ring and the waveguide. The coupling $\kext$ increases exponentially with decreasing $s$, whereas $\kint$ is independent of $s$. Yet, $\kint$ increases for smaller ring radii $R$ due to increasing bending loss \cite{vlasov_OE_bending_loss}. Smaller rings thus have wider linewidths. On the other hand, the spacing between the resonances shrinks with ring size since  $\FSR \propto R^{-1}$ \cite{bogaerts_LPR_ring_review}. An important figure of merit is the internal finesse $\Fint = \FSR/\kint$; the higher this finesse, the smaller the fraction of the resonances in the transmission spectrum. $\Fint$ is shown for rings with different widths and radii. Initially, $\Fint$ increases with $R$ for all widths since the reduction in $\kint$ exceeds the decrease in FSR. However, after some point the FSR decreases faster than the internal loss. The maximum $\Fint$ and the radius where it occurs depend on the width: the wider the ring, the higher the maximum internal finesse and the smaller the required radius. Wide rings thus allow for compact devices. In the remainder of this Letter a width of 1600 nm is used to achieve high $\Fint$ and single-mode operation.

\begin{figure}[tb]
\includegraphics{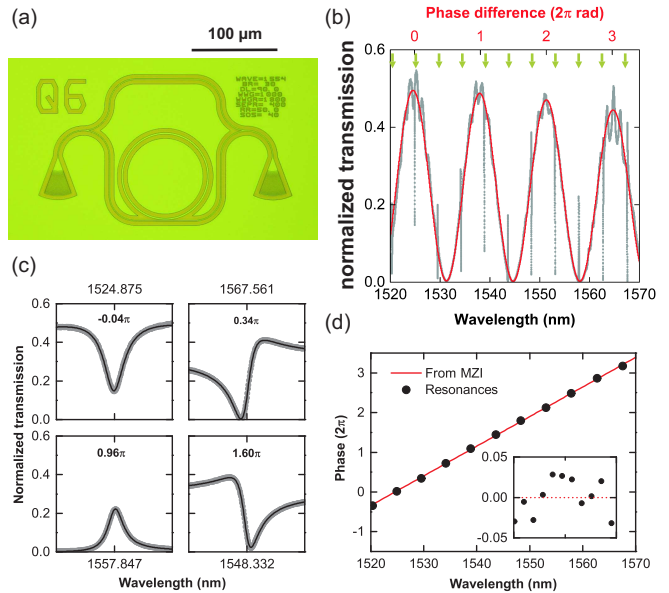}
\caption{(a) Optical micrograph of a MZI with a ring resonator.
(b) Transmission normalized to that of a calibration device. The resonances are indicated by the green arrows. The solid red curve is the fitted MZI response; from this the phase in the top axis is obtained.
(c) Zooms (100 pm span) of measured resonances with the best fitting response. The fitted value of $\phi$ is indicated. (d) Comparison between the phases obtained from the resonances and that from the MZI fringes. The inset shows the difference between the two.
\label{fig:MZI}}
\end{figure}

To illustrate the working principle of our method, the phase in a Mach-Zehnder interferometer (MZI) is determined using an integrated ring resonator and these values are verified using the wavelength-dependent fringes. A typical device is shown in Fig. \ref{fig:MZI}(a). It consists of input and output grating couplers with waveguides connecting to the interferometer which has a ring in its lower arm. The two arms of the MZI have different lengths. Such an unbalanced MZI displays sinusoidal fringes in its transmission [Fig. \ref{fig:MZI}(b)] since changing the wavelength adjusts the phase difference $\phi$. At the top of a fringe the two paths interfere constructively and $\phi =  0~(\mathrm{mod}~2\pi)$. On the other hand, destructive interference occurs at the bottom of a fringe where $\phi = \pi$. By fitting a sinusoidal function (with a non-flat overall transmission) to the fringes, $\phi$ is obtained for every wavelength as indicated at the top axis in Fig. \ref{fig:MZI}(b). Determining $\phi$ this way works well when a significant part of a full oscillation is visible in the transmission.

As explained above, a completely different method of finding $\phi$ is based on optical resonances. These appear in Fig. \ref{fig:MZI}(b) as narrow features. The radius of the 1600-nm-wide ring was $R=40 \un{\mu m}$ and $s= 600 \un{nm}$. Figure \ref{fig:rings}(c) indicates that such a resonator is overcoupled; the phase thus changes over the full $2\pi$ while sweeping over the narrow resonance. When zooming in onto some of the individual resonances, different peak shapes are observed as shown in Fig. \ref{fig:MZI}(c). This can be explained as follows: The output field is the sum of the fields from the lower and upper arm. The latter has encountered the ring and is thus multiplied by $h(\omega)$ [cf. Eq. (\ref{eq:transfer_ring})]. With the relative phase $\phi$, the transmission $T(\lambda)$ is:
\begin{equation}
T(\lambda) = T_0|e^{i\phi} + h(2\pi c/\lambda)|^2/4. \label{eq:MZI_transmission}
\end{equation}
Here, $T_0$ is the overall transmission and $c$ is the speed of light. Figure \ref{fig:MZI}(c) shows that with the indicated values for $\phi$, Eq. (\ref{eq:MZI_transmission}) fits the measured peak shapes well. For $\phi \approx 0$ the two paths interference constructively away from the resonance (where $\angle h = 0$). The phase change induced by the resonance reduces the constructive interference and the resonance appears as a dip. In contrast, for $\phi \approx \pi$ the off-resonant interference is destructive and the resonance is a peak. For other phases, the resonance shows increased and reduced transmission on different sides of the resonance wavelength as indicated by the other two panels in Fig. \ref{fig:MZI}(c). The phase extracted from the resonances is plotted versus wavelength in Fig. \ref{fig:MZI}(d). $\phi$ increases linearly and closely matches the phase independently extracted from the MZI fringes. Their differences (inset) have a mean value of $2\pi \times 3.7\times10^{-5}$ and a standard deviation of $2\pi \times 2.3\times10^{-2}$. These small values indicate the high accuracy of the extraction of the phase through the peak shapes.

\begin{figure*}[tb]
\includegraphics{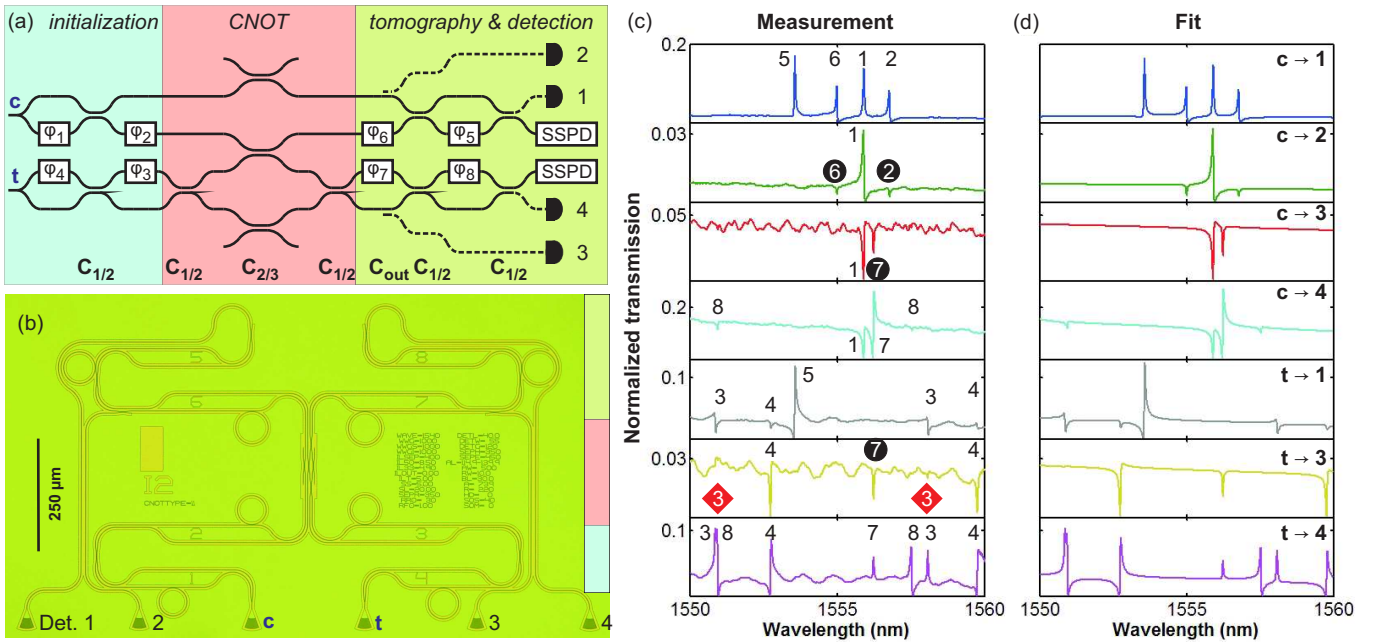}
\caption{(a) Schematic of the circuit for a CNOT operation with qubit initialization, tomography, and detection. The eight phase shifters and four monitor detectors (dashed) are indicated, as well as the power splitting ratios of the directional coupler. The top and bottom input represent the control (``c'') and target (``t'') qubit respectively. In the future, superconducting single photon detectors (SSPDs) will be used to detect single photons.
(b) Optical micrograph of a CNOT device with integrated ring resonators. The triangular structures are the grating couplers for coupling light on and off the chip using a fiber array.
(c) Measured transmissions through a CNOT device. The resonances are identified by the numbers; circles means that the resonance is only due to loss (i.e. not due to phase shifts), and diamonds highlights resonances that disappear exactly at $C_{1/2} = 1/2$.
(d) The best fit (``Fit 1'' in Table \ref{tab:fitresults}) to the data in panel (c).
\label{fig:CNOT}}
\end{figure*}

After this proof of principle, we apply our technique to a more complex photonic structure. Figure \ref{fig:CNOT}(a) shows a controlled-NOT circuit \cite{knill_nature_quantum_computation_linear_optics, kok_RMP_LOQC, ralph_PRA_CNOT_coincidence_theory} with circuitry for qubit initialization and tomography \cite{obrien_PRL_CNOT_tomography}. This device is made within our fully-integrated SiN platform for on-chip quantum-optics experiments \cite{poot_OE_quantum_circuits, schuck_natcomm_HOM, poot_apl_phaseshifter} and has two inputs and four monitor outputs; there are thus eight transmissions that can be measured. However, due to the waveguide layout [Fig. \ref{fig:CNOT}(a)] no light can go from input ``t'' to monitor port 2. With only 7 combinations is not possible to determine the 8 unknown phases in the circuit, let alone to find additional important parameters such as power splitting ratios of the directional couplers. However by using the phase shift of ring resonators a full characterization of this complex device is possible.

The measured transmissions shown in Fig. \ref{fig:CNOT}(c) display clear resonances. [Note that for better visibility a device with large $\kappa \sim 50 \un{pm}$ ($s = 450 \un{nm}$) is chosen; for rings with higher finesse the peak shape cannot be resolved visually in plots as Fig. \ref{fig:CNOT}(c).] First, the clearest resonances are found and each is fitted by a Fano function to accurately determine their position. Next, each resonances is attributed to a certain ring. The rings have different radii (25.5 to 29.0 $\un{\mu m}$) and thus a unique FSR.
Based on this, the remaining resonances are found. Fig. \ref{fig:CNOT}(c) shows for every resonance the attributed ring number. It is noted that some resonances appear as shallow dips (black circle). These are due to rings that do not interfere with light from a different path. The phase $\angle h$ does not matter in this case, the resonance only appears because light is lost via $\kint$ resulting in $|h(\omega_0)| < 1$. Finally, also the resonances of ring 3 are hardly visible in the $t\rightarrow 3$ trace. When $C_{1/2}$ is exactly equal to 1/2 these would not show up at all. Their small visibility thus indicates that $C_{1/2}$ is close its ideal value. However, determining this deviation requires a more quantitative analysis.

\begin{table}[!t]
\centering
\caption{Values and uncertainties of important fit parameters for the two different fits and the resulting root-mean-squared residuals (RMSE). Since $\phi_2$ and $\phi_6$ only occur in series [Fig. \ref{fig:CNOT}(a)], only their sum can be determined. The expected splitting ratios are indicated in the last column.}
\begin{tabular}{crrrc}
\hline
& Fit 1\hspace{7mm} & Fit 2\hspace{7mm} &  & Expected\\
\toprule
$C_{1/2}$           & 0.479 $\pm$ 0.005 & 0.492 $\pm$ 0.003 &  & 0.521 $\pm$ 0.022 \\
$C_{2/3}$           & 0.675 $\pm$ 0.004 & 0.670 $\pm$ 0.004 &  & 0.679 $\pm$ 0.019 \\
$\Cout$  & 0.031 $\pm$ 0.051 & 0.180 {(fixed)} \hspace{-1mm} &  & 0.180 $\pm$ 0.022 \\
$\phi_1$ &-2.360 $\pm$ 0.009 &-2.361 $\pm$ 0.009 & rad \\
$\phi_2 + \phi_6$   & 0.708 $\pm$ 0.011 & 0.722 $\pm$ 0.011 & rad \\
$\phi_3$ &-2.249 $\pm$ 0.026 &-2.300 $\pm$ 0.023 & rad \\
$\phi_4$ & 0.237 $\pm$ 0.044 & 0.287 $\pm$ 0.040 & rad \\
$\phi_5$ & 2.267 $\pm$ 0.025 & 2.326 $\pm$ 0.015 & rad \\
$\phi_7$ & 1.751 $\pm$ 0.033 & 1.839 $\pm$ 0.019 & rad \\
$\phi_8$ & 0.284 $\pm$ 0.016 & 0.279 $\pm$ 0.015 & rad \\
\hline
RMSE                & 0.0493\hspace{5mm} &  0.0494\hspace{5mm} &  \\
\hline
\end{tabular}
\label{tab:fitresults}
\end{table}

With the waveguide layout [Fig. \ref{fig:CNOT}(a)], the transmission profiles the circuit can be calculated \cite{poot_OE_quantum_circuits} and fitted to the measured traces by varying the a-priory unknown parameters (e.g. the phases $\phi_i$ and the coupling ratios $C_i$). Due to the large number of fit parameters (44) it is important to find good starting values for the fit. Another aspect of fitting seven spectra simultaneously is how these are weighted. Different combinations of input and output can have widely different transmissions, depending on the parameters of the circuit. Hence, a combination with low transmission may not be accurately represented in an unweighted fit since the fit minimizes the squared difference between measured and calculated spectra. To prevent this, the profiles are weighted using the inverse of the outputs averaged over all phases that are calculated using the model for the circuit.

As shown in Fig. \ref{fig:CNOT}(d) the fits accurately reproduce all features observed in the data, including the magnitudes and peak shapes. The resulting parameters of interest are given in Table \ref{tab:fitresults}; the details of the fit and its results can be found in the Supplementary Information. The phases $\phi_i$ are found with small uncertainties. The same holds for the coupling ratios $C_{1/2}$ and $C_{2/3}$. Their fitted values are close those expected from measurements on directional coupler with varying lengths (last column) \cite{poot_OE_quantum_circuits}. They are also close to the ideal values of 1/2 and 2/3 respectively, highlighting our control over the fabrication process. $\Cout$ is, however, lower than expected and has a larger uncertainty. A second fit with $\Cout$ fixed to its expected value is also performed. In this case, the RMSE hardly increases and also the other parameters only change slightly. Using the phase shifts induced by the rings we have thus extracted the important parameters

Finally, we note that the results presented here are not limited to ring resonators. Importantly, it is also possible to use spectrally isolated resonances, for example using defect cavities in photonic crystal structures \cite{joannopoulos_book_PhC_molding}. This way, it is possible to integrated many resonances in spectral ranges where the circuit operation is not perturbed. The resonant phase shifts employed here thus provide a powerful method to characterize linear optical circuits.

\section*{Supplementary Material}
Details about the fitting procedure for the CNOT device and values of all fit parameters can be found in the Supplementary Material.

\begin{acknowledgments}
This work was partly funded by the Packard Foundation. H.X.T. acknowledges support from a career award from National Science Foundation. Facilities used were supported by Yale Institute for Nanoscience and Quantum Engineering and NSF MRSEC DMR 1119826.
\end{acknowledgments}


%

\appendix
\renewcommand{\thetable}{S\arabic{table}}%
\renewcommand{\thefigure}{S\arabic{figure}}%
\renewcommand{\theequation}{S\arabic{equation}}%

\clearpage
\section*{Supplementary Material}
This Supplementary Information gives details about the procedure used to extract the phases of the CNOT circuit. First, resonances (i.e. regions with large curvature) are identified in the seven transmission profiles [Fig. 3(c)], and their exact frequency is determined by fitting a Fano function to the resonance. Since the rings have different radii, their FSRs are different. In an iterative procedure, each resonance is attributed to a particular ring. This way, one also finds the FSR of each ring as given in Table \ref{tab:ringdata}. The FSR clearly decreases with increasing $R$. However, the group index $n_g = \lambda^2/(2\pi R~\mathrm{FSR})$ is independent of $R$.

\begin{table}[htbp]
  \centering
  \caption{Radii $R$, free spectral range FSR (mean and standard deviation over the different transmission traces), and group index $n_g$ of the eight rings. 
  }
    \begin{tabular}{cccccc}
    \hline
    \bf{Ring}  &  $ {\mathbf{ R}}$ \bf{($\mu$m)} & \bf{FSR (nm)} & \hspace{5mm} & ${\mathbf n_g}$ \\
    \toprule
    1     &        25.5  &     7.5102  $\pm$     0.0006  &  &   2.0069  \\
    2     &        26.0  &     7.3602  $\pm$     0.0012  &  &   2.0084   \\
    3     &        26.5  &     7.2281  $\pm$     0.0038  &  &   2.0066    \\
    4     &        27.0  &     7.0837  $\pm$     0.0093  &  &   2.0095   \\
    5     &        27.5  &     6.9520  $\pm$     0.0029  &  &   2.0104   \\
    6     &        28.0  &     6.8358  $\pm$     0.0046  &  &   2.0080  \\
    7     &        28.5  &     6.7181  $\pm$     0.0026  &  &   2.0074  \\
    8     &        29.0  &     6.6038  $\pm$     0.0039  &  &   2.0069   \\
    \hline
    \end{tabular}
  \label{tab:ringdata}
\end{table}

For constant $n_g$ the resonance \emph{frequencies} of a ring are equally spaced. However, since the experimental traces are measured versus wavelength $\lambda$, it is convenient to introduce a new variable $x(\lambda) = \lambda_{\mathrm{ref}} - \lambda_{\mathrm{ref}}^2/\lambda$. This way, $x$ has the dimension of length and can be directly compared to wavelengths. In particular, one has $x(\lambda_{\mathrm{ref}}+\Delta\lambda) \approx \Delta \lambda$ to first order in $\Delta \lambda$. Yet, with this definition the resonance positions of ring $i$, $x_0^{(i)}$, are equally spaced in $x$ by the free spectral range $\mathrm{FSR}_i$. In the following, we choose the reference wavelength $\lambda_{\mathrm{ref}} = 1550 \un{nm}$.

The transmission through the entire device is calculated by cascading scattering matrices of all the different elements. This is repeated for all experimental wavelengths, and the fields at the outputs are squared to obtained the optical power. When measuring, it is possible that the position of the fiber array is not exactly above the grating couplers of the device. To account for small differences in the absolute power, adjustments factors for the transmission of the four monitor ports $M_i$ are introduced. If all $M_i=1$, the absolute transmissions would perfectly match those of the calibration devices. However, as Table \ref{tab:fitextened} shows, the values of $M_i$ are slightly below one, indicating a small deviation from the optimal position. A similar adjustment is made for the ratio $R$ between the power going into port 'c' compared to port 't'.

Since the span of the data shown in Fig. 3(d) exceeds the FSR of the rings, multiple resonances of a single ring can appear. Equation (1) is for an isolated resonance only. However, ring resonators have a transfer function that is equal to that of an etalon:
\begin{equation}
h(x) = \frac{~r-\rho \exp\{-2\pi i (x-x_0)/\FSR\}}{1-r\rho \exp\{-2\pi i (x-x_0)/\FSR\}}. \label{eq:etalon}
\end{equation}
Here, $r$ is the fraction of the light field in the input waveguide that does not enter the ring at all, and $\rho$ is the fraction that remains after a round trip. By comparing Eq. (\ref{eq:etalon}) with Eq. (1) in the limit of small loss (i.e, $1-r, 1-\rho \ll 1$), relations are found between $r$ and $\rho$, and $\kint$ and $\kext$: The coupling linewidth is $\kext \approx \FSR (1 - r)/\pi$, whereas the internal loss $\kint \approx  \FSR (1- \rho)/\pi$. As an example, consider ring 1, which has $\FSR = 7.5102 \un{nm}$, $r = 0.9833$, and $\rho = 0.9953$ according to Tables \ref{tab:ringdata} and \ref{tab:fitextened}. This gives $\kint = 11 \un{pm}$ and $\kext = 39 \un{pm}$, which are in good agreement with the data shown in Fig. 1(c). The internal finesse is $\Fint = \FSR/\kint \approx \pi/(1-\rho) = 6.7\times 10^2$.

The fit in Fig. 3(d) has 44 free parameters. Table \ref{tab:fitextened} shows their values and uncertainties for the two fits. The first 10 parameters are the phases and coupling ratios that we want to determine using our fitting procedure. The second group (11..20) contains correction factors for the absolute transmission through the device, whereas the third group (21..44) are the properties of the eight rings. As noted before, $\Cout$ cannot be determined accurately. This is reflected by its uncertainty, which is an order of magnitude larger than those
in $C_{1/2}$ and $C_{2/3}$. The power measured coming from output ports 2 (3) is proportional to $M_{2(3)}\Cout$. The large uncertainty in $\Cout$ thus also gives a large fit uncertainty in $M_{2,3}$, as well as in the corresponding slopes. For the second fit, the value of $\Cout$ is fixed to the expected value of 0.18 and the uncertainties in $M_{2,3}$ are greatly reduced.

\begin{table*}[htbp]
  \centering
  \caption{Values and uncertainties of all fit parameters for the two different fits. The fits correspond to Table 1 and Fig. 3(d) in the main text. The fit parameters are explained in the main and supplementary text.}
    \begin{tabular}{|l|lc|rr|rrr|}
    \toprule
    \bf{No.}   & \bf{Parameter}      &  \bf{Symbol} & {\bf Value 1} & {\bf Unc. 1} & {\bf Value 2} & {\bf Unc. 2} &  \\
    \hline
1               & Phase 1         & $\phi_1$        & -2.359          & 0.009           & -2.359          & 0.009           & rad \\
2               & Phase 2  + Phase 6 & $\phi_2 + \phi_6$ & 0.709           & 0.011           & 0.722           & 0.011           & rad \\
3               & Phase 3         & $\phi_3$        & -2.252          & 0.026           & -2.301          & 0.023           & rad \\
4               & Phase 4         & $\phi_4$        & 0.240           & 0.044           & 0.289           & 0.040           & rad \\
5               & Phase 5         & $\phi_5$        & 2.270           & 0.025           & 2.327           & 0.015           & rad \\
6               & Phase 7         & $\phi_7$        & 1.753           & 0.033           & 1.839           & 0.019           & rad \\
7               & Phase 8         & $\phi_8$        & 0.286           & 0.016           & 0.281           & 0.015           & rad \\
8               & Splitting ratio 50/50 coupler & $C_{1/2}$       & 0.479           & 0.005           & 0.492           & 0.003           &  \\
9               & Splitting ratio 67/33 coupler & $C_{2/3}$       & 0.674           & 0.004           & 0.670           & 0.004           &  \\
10              & Splitting ratio monitor coupler & $\Cout$ & 0.035           & 0.051           & 0.180           & fixed           &  \\
                &                 &                 &                 &                 &                 &                 &  \\
11              & Transm. monitor port 1    & $M_1$           & 0.782           & 0.037           & 0.878           & 0.025           &  \\
12              & Transm. monitor port 2    & $M_2$           & 5.0           & 7.4           & 0.957           & 0.024           &  \\
13              & Transm. monitor port 3    & $M_3$           & 4.5           & 6.6           & 0.849           & 0.008           &  \\
14              & Transm. monitor port 4    & $M_4$           & 0.673           & 0.027           & 0.761           & 0.016           &  \\
15              & Transm. ratio "t" to "c"  & $R$             & 0.592           & 0.011           & 0.603           & 0.011           &  \\
16              & Slope port 1    & -$\pderl{M_1}{x}$ & 0.019           & 0.002           & 0.0210          & 0.0021          &  \\
17              & Slope port 2    & -$\pderl{M_2}{x}$ & 0.2             & 0.3             & 0.0337          & 0.0016          &  \\
18              & Slope port 3    & -$\pderl{M_3}{x}$ & 0.05            & 0.07            & 0.0087          & 0.0003          &  \\
19              & Slope port 4    & -$\pderl{M_4}{x}$ & 0.0161          & 0.0007          & 0.0183          & 0.0005          &  \\
20              & Slope ratio     & -$\pderl{R}{x}$  & 0.0046          & 0.0003          & 0.0046          & 0.0003          &  \\
                &                 &                 &                 &                 &                 &                 &  \\
21              & Resonance position ring 1 & $x_0^{(1)}$     & 5.8921          & 0.0004          & 5.8922          & 0.0004          & nm \\
22              & Coupling factor ring 1    & $r^{(1)}$       & 0.9833          & 0.0002          & 0.9833          & 0.0002          &  \\
23              & Round trip factor ring 1  & $\rho^{(1)}$    & 0.9963          & 0.0003          & 0.9963          & 0.0002          &  \\
24              & Resonance position ring 2 & $x_0^{(2)}$     & 6.7553          & 0.0013          & 6.7552          & 0.0013          & nm \\
25              & Coupling factor ring 2    & $r^{(2)}$       & 0.9850          & 0.0010          & 0.9850          & 0.0010          &  \\
26              & Round trip factor ring 2  & $\rho^{(2)}$    & 0.9971          & 0.0005          & 0.9972          & 0.0005          &  \\
27              & Resonance position ring 3 & $x_0^{(3)}$     & 0.8717          & 0.0019          & 0.8717          & 0.0018          & nm \\
28              & Coupling factor ring 3    & $r^{(3)}$       & 0.9811          & 0.0013          & 0.9832          & 0.0012          &  \\
29              & Round trip factor ring 3  & $\rho^{(3)}$    & 0.9962          & 0.0013          & 0.9954          & 0.0013          &  \\
30              & Resonance position ring 4 & $x_0^{(4)}$     & 2.7486          & 0.0006          & 2.7488          & 0.0006          & nm \\
31              & Coupling factor ring 4    & $r^{(4)}$       & 0.9814          & 0.0005          & 0.9821          & 0.0005          &  \\
32              & Round trip factor ring 4  & $\rho^{(4)}$    & 0.9968          & 0.0006          & 0.9957          & 0.0005          &  \\
33              & Resonance position ring 5 & $x_0^{(5)}$     & 3.5649          & 0.0007          & 3.5647          & 0.0007          & nm \\
34              & Coupling factor ring 5    & $r^{(5)}$       & 0.9833          & 0.0007          & 0.9846          & 0.0005          &  \\
35              & Round trip factor ring 5  & $\rho^{(5)}$    & 0.9966          & 0.0007          & 0.9953          & 0.0005          &  \\
36              & Resonance position ring 6 & $x_0^{(6}$      & 4.9919          & 0.0012          & 4.9919          & 0.0012          & nm \\
37              & Coupling factor ring 6    & $r^{(6)}$       & 0.9842          & 0.0009          & 0.9842          & 0.0009          &  \\
38              & Round trip factor ring 6  & $\rho^{(6)}$    & 0.9963          & 0.0005          & 0.9963          & 0.0005          &  \\
39              & Resonance position ring 7 & $x_0^{(7)}$     & 6.2174          & 0.0004          & 6.2175          & 0.0004          & nm \\
40              & Coupling factor ring 7    & $r^{(7)}$       & 0.9851          & 0.0004          & 0.9852          & 0.0004          &  \\
41              & Round trip factor ring 7  & $\rho^{(7)}$    & 0.9971          & 0.0001          & 0.9970          & 0.0001          &  \\
42              & Resonance position ring 8 & $x_0^{(8)}$     & 0.9512          & 0.0009          & 0.9508          & 0.0009          & nm \\
43              & Coupling factor ring 8    & $r^{(8)}$       & 0.9842          & 0.0008          & 0.9844          & 0.0008          &  \\
44              & Round trip factor ring 8  & $\rho^{(8)}$    & 0.9968          & 0.0002          & 0.9968          & 0.0002          &  \\
    \hline
                & Root-mean-squared error   & RMSE            & 0.0493 &       & 0.0494 &       &  \\
    \hline
    \end{tabular}
  \label{tab:fitextened}
\end{table*}

\end{document}